\documentstyle[aps,twocolumn,epsf]{revtex}
\begin{document}
\draft
\title{Dynamic wormholes}
\author{Sean A. Hayward}
\address{Yukawa Institute for Theoretical Physics, Kyoto University,
Kitashirakawa, Sakyo-ku, Kyoto 606-8502, Japan\\
\tt hayward@yukawa.kyoto-u.ac.jp}
\date{Revised 20th January 1999}
\maketitle

\begin{abstract}
A new framework is proposed for general dynamic wormholes,
unifying them with black holes.
Both are generically defined locally by outer trapping horizons,
temporal for wormholes and spatial or null for black and white holes.
Thus wormhole horizons are two-way traversible,
while black-hole and white-hole horizons are only one-way traversible.
It follows from the Einstein equation that 
the null energy condition is violated everywhere on a generic wormhole horizon. 
It is suggested that quantum inequalities constraining negative energy 
break down at such horizons. 
Wormhole dynamics can be developed as for black-hole dynamics,
including a reversed second law 
and a first law involving a definition of wormhole surface gravity.
Since the causal nature of a horizon can change,
being spatial under positive energy 
and temporal under sufficient negative energy,
black holes and wormholes are interconvertible.
In particular, if a wormhole's negative-energy source fails,
it may collapse into a black hole.
Conversely, irradiating a black-hole horizon with negative energy
could convert it into a wormhole horizon.
This also suggests a possible final state of black-hole evaporation:
a stationary wormhole.
The new framework allows a fully dynamical description of 
the operation of a wormhole for practical transport,
including the back-reaction of the transported matter on the wormhole.
As an example of a matter model, 
a Klein-Gordon field with negative gravitational coupling 
is a source for a static wormhole of Morris \& Thorne.
\end{abstract}
Space-time wormholes, short cuts between otherwise distant 
or even unconnected regions of the universe,
have become a common feature in science fiction.
That they might be science fact is suggested by 
the simplest stationary black-hole solutions to the Einstein equations,
which do indeed have the global structure of wormholes,
albeit just barely failing to be traversible.
Morris \& Thorne\cite{MT} revived interest in the possibility of wormholes 
for practical travel,
showing that it is theoretically possible in Einstein gravity, 
provided sufficient negative energy can be maintained.
Although mundane matter has positive energy,
quantum theory predicts negative energy during particle creation,
such as in the Casimir effect.
Morris \& Thorne considered the simplest case 
of static, spherically symmetric wormholes,
and subsequently there has been some controversy 
about whether negative energy is necessary in general,
or if so, whether it can be avoided by travellers navigating the wormhole.
This is discussed in detail by Hochberg \& Visser\cite{HV1,HV2},
who independently reach the conclusion (confirmed in this article) that 
negative energy is required generically.

The fundamental problem is that 
there is no agreed idea of what wormholes actually are in general.
This letter proposes such a definition and discusses widespread consequences.
The concepts involved are mostly familiar in relativity,
e.g.\ Hawking \& Ellis\cite{HE},
and otherwise defined in the author's previous papers 
on black holes\cite{H1,H2}, 
so will not be explained in detail here.
Similarly, proofs will be avoided since 
they are simple extensions or converses of previous results for black holes
and will anyway be spelled out in a longer article\cite{H3}.

A static wormhole may be defined simply by 
a minimal surface in a static spatial (spacelike) hypersurface,
so that it is locally the narrowest section of a tunnel 
between two larger regions.
Morris \& Thorne gave further wormhole criteria 
which will not be considered here,
except for the basic context of Einstein gravity.
This effectively includes other gravitational theories like Brans-Dicke
which involve what can be treated as an Einstein equation with other fields,
as emphasised by Hochberg \& Visser.

To define dynamic wormholes, an analogous idea (refined below) is that 
the spatial surface is minimal in some spatial hypersurface.
This implies that the surface is generically
a (future or past) trapped surface,
usually associated with black or white holes.
Here genericity means stability in the sense that 
any sufficiently small variation of a trapped surface is also a trapped surface,
and therefore also minimal in some spatial hypersurface.\footnote
{The non-generic case is a surface which is doubly marginal somewhere:
arbitrary variations of such a surface are usually not minimal
in any spatial hypersurface.
This class includes static minimal surfaces 
and suggests a minor generalisation from static to stationary wormholes:
a stationary outer trapping horizon.}
Thus a wormhole must generically be a region of space-time.
This differs qualitatively from a static wormhole, 
which is the temporal (timelike) hypersurface generated by the minimal surfaces.
For a wormhole region to be two-way traversible,
it should have two boundaries, 
referred to here as wormhole horizons and elsewhere as wormhole throats,
which are in causal contact.
Local two-way traversibility suggests that 
these horizons should be temporal hypersurfaces. 
Moreover, one expects the horizons to be foliated by marginal surfaces 
(marginally trapped surfaces), 
as occurs for black-hole apparent horizons according to Hawking \& Ellis.
Proving this is non-trivial\cite{KH},
so it seems more practical to assume it as part of the definition, as follows.

A hypersurface foliated by marginal surfaces will be called a 
{\sl trapping horizon}, following Ref.~\cite{H1}.
The marginal surfaces are a dynamic generalisation of 
the static minimal surfaces,
except for the so-called flare-out condition 
selecting minimal surfaces from extremal ones.
For the generic flare-out condition, 
this letter proposes an {\sl outer}\/ trapping horizon 
as defined in Ref.~\cite{H1}.
This is a strict inequality, 
excluding cases of equality which describe degenerate wormholes, 
for reasons explained below.
It reduces in the static case to the strict static flare-out condition.
In short, a generic wormhole horizon is defined as a
{\sl temporal outer trapping horizon}.

The definition implies that the marginal surfaces are minimal 
in the null direction of marginality.
Thus each surface is locally the narrowest section of a null hypersurface,
naturally generalising the static definition.
This sort of idea was first suggested by Page, according to Morris \& Thorne.
A precise formulation, similar to that proposed here, 
is suggested independently by Hochberg \& Visser,
and indeed their strict flare-out condition is implied by the above definition.
However, the converse is not so, 
since Hochberg \& Visser specifically allow spatial horizons,
in which case their flare-out condition 
selects maximal rather than minimal surfaces in a time-symmetric hypersurface.

An immediate consequence of the definition and the Einstein equation is that 
{\sl the null energy condition is violated 
everywhere on a generic wormhole horizon}\cite{H3}.
This negative-energy theorem offers a clean resolution to the above controversy.
This simple result gives more information than 
the topological censorship theorem of Friedman et al.\cite{FSW}, 
which showed that 
the null energy condition excludes wormholes as defined globally,
in asymptotically flat space-times.
Such global assumptions have been avoided here in favour of 
a local, dynamical definition of wormhole horizon.
Similarly, the negative-energy property is purely local,
applying to each point of the wormhole horizon.
Thus the definition could be relaxed so that 
the horizon need be temporal only somewhere,
in which case only those parts need have negative energy.

The definition could also be relaxed to allow 
degenerate outer trapping horizons, 
i.e.\ with the flare-out inequality becoming non-strict, in which case 
negative energy could just barely be avoided in very special cases.
However, such wormholes are therefore unstable 
against arbitrarily small variations due to positive-energy matter,
such as an explorer attempting to traverse the wormhole.
Such traversibly unstable wormholes are useless for practical transport,
so there is little lost by excluding them from the definition,
which simplifies matters somewhat.
Including them leads to more complicated negative-energy results,
comparable to those of Hochberg \& Visser.

Ref.~\cite{H1} actually defined black (respectively white) holes by
future (respectively past) outer trapping horizons.
However, this was intended to apply under the usual positive-energy conditions,
under which outer trapping horizons are spatial or null,
reflecting the intuitive idea that one may enter but not leave a black hole,
and leave but not enter a white hole.
The above negative-energy theorem is a converse of this result.
Thus the framework actually unifies black holes and wormholes,
revealing that they are similar objects, geometrically and physically.
Indeed, {\sl black holes and wormholes are locally identical 
except for the causal nature of the horizons.}
If the horizon is temporal and therefore two-way traversible,
we can call it a wormhole horizon,
while if it is spatial or null and therefore only one-way traversible,
we can call it a black-hole or white-hole horizon,
respectively if the trapped region is to the future or past.
One may refer to all collectively simply as space-time holes.

This connection between black holes and wormholes 
seems to have gone unnoticed elsewhere.
Indeed, even the most seminal authors\cite{MT,HV1} have instead stressed that 
wormholes are quite different from black holes, 
sternly warning against comparisons.
This seems to be due to misconceptions engendered by 
the old global paradigm for black holes in terms of event horizons.
The connection arises instead from 
the new local, dynamical paradigm for black holes 
in terms of trapping horizons\cite{H1,H2}.

This theory of black-hole dynamics induces a corresponding theory of 
{\sl wormhole dynamics}, as will be detailed in Ref.~\cite{H3}.
Some basic laws of wormhole dynamics can be obtained simply 
by adapting those for black and white holes.
In particular, the second law of black-hole dynamics of Ref.~\cite{H1}
becomes reversed under negative energy, 
yielding a {\sl second law}\/ of wormhole dynamics:
future, past or stationary wormhole horizons 
respectively have decreasing, increasing or constant area.

In spherical symmetry, the definition of {\sl surface gravity}\/
proposed in Ref.~\cite{H2} for dynamic black or white holes 
may also be applied to wormholes,
providing a physical measure of the strength of the wormhole.
This seems to have been lacking even for static wormholes,
for which one finds that the surface gravity over circumference equals
tension minus energy density.
Positivity of the surface gravity is equivalent to 
the strict flare-out condition, giving the latter a physical meaning.
The non-strict flare-out condition would allow vanishing surface gravity,
describing degenerate wormholes.

The surface gravity occurs in a {\sl first law}\/ of wormhole dynamics
with the same form as the first law of black-hole dynamics of Ref.~\cite{H2}.
The restriction to spherical symmetry will be removed 
in forthcoming work\cite{MH}.
If quantum temperature and gravitational entropy are related to 
wormhole surface gravity and area by a generalisation of 
the (partly conjectural) relationship for stationary black holes,
then we would have another new field, wormhole thermodynamics.

\begin{figure}
\centerline{\epsfxsize=5cm \epsfbox{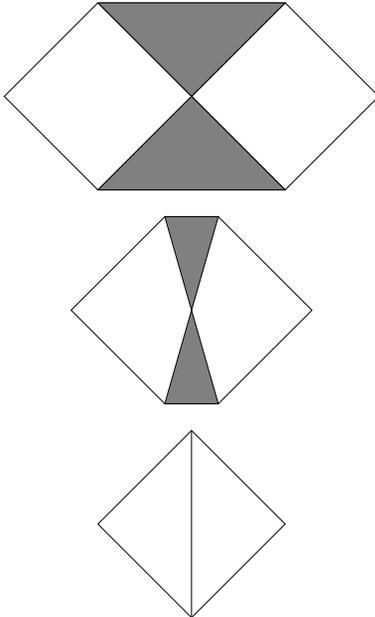}}
\caption{Penrose diagrams (see e.g.\ Hawking \& Ellis) of 
(i) the Schwarzschild black-hole space-time,
(ii) a modification to a dynamic wormhole,
and (iii) a further modification to a static wormhole.
Shading indicates trapped regions.}
\label{ex}
\end{figure}

Some spherically symmetric examples are as follows.
Fig.\ref{ex} depicts Penrose diagrams of 
(i) the Schwarzschild space-time, 
the unique vacuum black-hole solution, which is static;
(ii) a modification in which the trapped regions have shrunk spatially
so that the trapping horizons become temporal;
and (iii) a further modification in which the trapped regions have shrunk 
away completely and the trapping horizons have coalesced,
yielding a static Morris-Thorne wormhole.
Then (ii) is a dynamic wormhole by either the current definition 
or the global definition that an observer may pass from one asymptotic region 
to the other in either direction.
This example illustrates that 
wormholes may contain either future or past trapped surfaces,
the boundaries being either future or past trapping horizons.

The dynamics of a trapping horizon actually depend on 
only two components of the Einstein equation,
so it is not difficult to develop intuition 
about the qualitative behaviour of the horizon.
The main factor is the effective energy density 
in the null direction of marginality, 
generally including a shear term which acts like positive energy.
If this effective energy is positive, zero or negative, 
an outer trapping horizon is respectively spatial, null or temporal,
and as the effective energy strengthens or weakens,
the horizon bends respectively away from or towards 
the null direction of marginality.
This suffices to understand the various examples given below.

In particular, it should be stressed that 
the behaviour of wormholes and black holes as described below 
is not speculation (except where indicated) 
but mathematical consequences of the definitions and Einstein theory,
verifiable by the methods of Ref.\cite{H1}.
For example, 
a stationary wormhole consists of the coincidence of two trapping horizons,
so it will bifurcate under a generic non-stationary variation.
The above methods indicate that this forms future or past trapping horizons 
if the effective energy is weakening or strengthening respectively.

\begin{figure}
\centerline{\epsfxsize=6cm \epsfbox{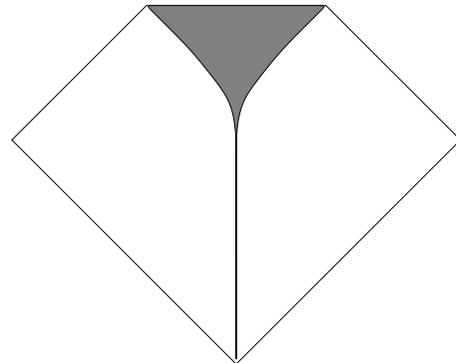}}
\caption{If a wormhole's negative-energy source fails, it becomes a black hole.
An initially static trapping horizon like Fig.\ref{ex}(iii) 
bifurcates and becomes dynamic as in Fig.\ref{ex}(ii),
then null as in Fig.\ref{ex}(i).}
\label{wb}
\end{figure}

A unified treatment of black holes and wormholes 
is not merely useful but logically necessary,
since a trapping horizon may change its causal nature as it develops,
being spatial under positive energy 
and temporal under sufficient negative energy.
A practical way to make this point is:
what happens to a wormhole if its negative-energy generator fails?
The answer is that it becomes a black or white hole, at least locally.
This can be seen as follows, fixing spherical symmetry.
Suppose that a static wormhole like (iii) has been constructed,
necessarily maintained by negative energy,
and that the negative energy subsequently weakens.
Then the horizon generally bifurcates, opening up a future trapped region 
and forming a dynamic wormhole locally like (ii).
If the energy disperses smoothly and completely to leave vacuum, 
the solution must become Schwarzschild like (i), 
the horizons smoothly becoming null as in Fig.\ref{wb}.
Thus the wormhole has a collapsed into a black hole. 

A physical interpretation is that 
{\sl stationary black holes are ground states for wormholes.}
In this view, wormholes are negative-energy excited states,
which relax to a ground state in the absence of the excitation energy.
Since stationary black holes are a physically important limit of wormholes,
it is useful to have a framework like that of Ref.~\cite{H1} 
which automatically includes this limit.
The Hochberg-Visser approach differs in this respect,
their suggested measure of flare-out vanishing for stationary black holes.

Alternatively, if some positive-energy matter subsequently crosses the horizon, 
it becomes spatial, corresponding to a dynamic black hole.
Then again, 
if starfleet engineers succeed in repairing the negative-energy generator,
the horizon becomes temporal again and the wormhole can be restored.
In this case, the wormhole has become a black hole and then a wormhole again,
by any local reckoning.
This makes it clear that wormhole physics cannot ignore black holes.
Any actual wormhole would be in danger of becoming a black hole 
if the negative-energy source failed,
or was overwhelmed by normal positive-energy matter.
An anti-wormhole weapon could therefore be made 
of any sufficiently concentrated mass.

\begin{figure}
\centerline{\epsfxsize=6cm \epsfbox{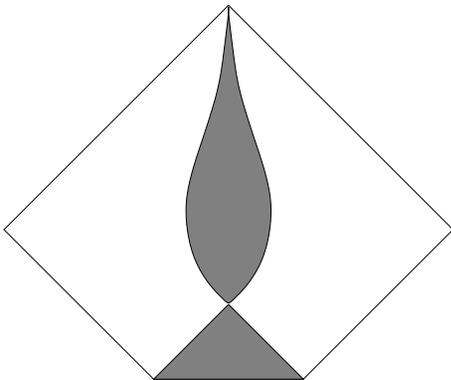}}
\caption{Conversion of a black hole into a wormhole 
by irradiation with negative energy.
An initially Schwarzschild black hole as in Fig.\ref{ex}(i)
becomes a dynamic wormhole as in Fig.\ref{ex}(ii).
As depicted, the wormhole may eventually settle down to a static state 
as in Fig.\ref{ex}(iii).
The negative-energy material might be just Hawking radiation, suggesting that 
the final state of black-hole evaporation might be a stationary wormhole.}
\label{bw}
\end{figure}

Conversely, if negative energy is appreciable,
black-hole physics cannot ignore wormholes.
Irradiating a black-hole horizon with negative energy 
would convert it into a wormhole horizon as in Fig.\ref{bw}.
This might be a practical way to make a wormhole,
though astrophysical black holes, presumably formed by gravitational collapse,
are not expected to have the right topology for global traversibility.
Black holes formed by quantum fluctuations, 
either in the early universe or by some hypothetical quantum engineering, 
may well have wormhole topology,
judging by the simplest stationary solutions.
This at least suggests further science fiction plots.
Another such plot stems from the fact that 
an explorer falling into a black hole may not be trapped after all,
but could escape if rescuers could beam enough negative energy into the hole 
quickly enough.

A theoretical example of black-hole-to-wormhole conversion already exists:  
black-hole evaporation.
In this case, the negative-energy material is just the Hawking radiation 
in semi-classical theory, 
which is expected to cause the trapping horizons to become temporal and shrink.
Locally these are future wormhole horizons.
The final state of black-hole evaporation 
has been the subject of much discussion,
to which another possibility is now suggested: 
an initially stationary black hole (of wormhole topology like Schwarzschild) 
might evaporate to leave a finally stationary wormhole.
This would be possible if the particle production decreases 
as the trapping horizon shrinks,
allowing the two horizons to approach slowly and asymptotically coalesce 
as in Fig.\ref{bw}.
The existence of such a final state requires a stationary wormhole 
which is a self-consistent semi-classical solution,
with particle production on the wormhole background
providing the negative-energy matter supporting the wormhole.
Such solutions do indeed exist, according to Hochberg et al.\cite{HPS}.
Hochberg \& Kephart\cite{HK} also mentioned the possibility of
wormholes formed by black-hole evaporation.
This might also resolve the black-hole information puzzle,
since the trapped region shrinks to nothing 
and its contents must therefore re-emerge.

A practical problem stems from the fact that
the actual use of a wormhole for transport would involve 
mundane positive-energy matter traversing it:
too much such transport would convert the wormhole into a black hole.
Keeping the wormhole viable, defying its natural fate as a black hole,
would require additional negative energy to balance the transported matter.
Previous work has been unable to deal with this consistently,
instead ignoring the back-reaction of the transported matter on the wormhole.
The new framework allows a consistent description of wormhole operation, 
sketched as follows.

\begin{figure}
\centerline{\epsfxsize=5cm \epsfbox{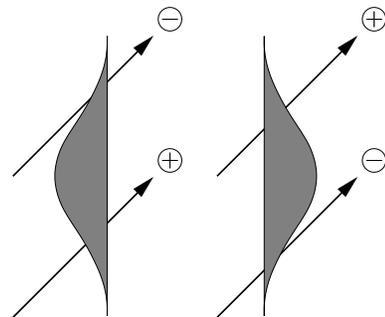}}
\caption{Operation of an initially static wormhole 
to transport positive-energy matter,
balanced by additional negative energy to return it to a static state.
The wormhole horizon bifurcates, opening up a region which is
 future trapped if the positive energy is sent first and
past trapped if it is sent last.
The diagrams illustrate the simplest case of pure radiation.}
\label{op}
\end{figure}

During operation, an initially stationary wormhole horizon would bifurcate,
opening up a trapped region, but could be closed up to a stationary state again 
by a careful balance of positive and negative energy, as in Fig.\ref{op}.
The temporary wormhole region could be either future or past trapped,
depending on whether the positive-energy matter is sent in before or after 
the compensating negative energy, respectively.
After operation, the wormhole would be respectively smaller or larger.
Consequently an alternating process is suggested,
sandwiching the transported matter between bursts of negative energy,
so that the wormhole is kept tight when not in use and dilated as necessary.

If the negative-energy source fails during the dynamical stage,
the wormhole would become locally a black hole or white hole, respectively.
In the latter case, 
the white-hole horizons would presumably collapse towards each other,
meet and pass through each other as in the Schwarzschild space-time,
leaving a black hole again.
Either way, wormhole experiments would risk accidental creation of black holes,
subsequently littering the universe.

The above analysis reveals that 
an operating wormhole would be continually pulsating,
a picture of dynamic wormhole operation that seems to be new.
Although the example was implicitly spherically symmetric, it seems likely that 
a more practical wormhole design would be otherwise, say axially symmetric, 
with the transported matter traversing the equatorial plane 
and the negative energy being pumped through the poles.

The strict nature of the negative-energy theorem indicates 
another practical problem of wormhole design,
that the total energy including the transported matter must be negative.
Morris \& Thorne, quoting Sagan, suggested that 
the transported matter might be somehow shielded from the negative energy,
but such a region of positive energy cannot cross a wormhole horizon,
even allowing the more relaxed definition of a partially temporal horizon.
Negative energy is required at precisely those points of the horizon 
which are locally two-way traversible.
Consequently it seems that wormhole transport requires negative-energy matter 
which couples so weakly to mundane matter that it has no harmful effects.
The use of a wormhole merely for signalling would involve weaker requirements.

The nature of the negative-energy matter has been left unspecified above,
in order to obtain results which are as general as possible.
Two main sources of negative energy have been considered in the literature:
alternative theories of gravity and particle creation in quantum field theory.
In the latter case, in flat space-time, 
Ford \& Roman\cite{FR1,FR2} derived so-called quantum inequalities 
constraining the negative energy, for instance that 
its magnitude is less than the Planck constant divided by the time it persists,
and argued that this would confine wormholes by the Planck scale.
However, when Pfenning \& Ford\cite{PF} generalized this method 
to static space-times, 
the inequalities for static observers were found to depend singularly 
on the norm of the static Killing vector, 
which physically encodes the gravitational redshift.
For instance, for a Schwarzschild black hole, 
the inequalities break down at the horizons\cite{PF,FT}.
It has been argued that the flat-space quantum inequalities still hold 
for certain timescales\cite{PF,FPR},
but these timescales vanish at the horizons.
Thus it may be conjectured that 
quantum inequalities generally break down at trapping horizons.
For instance, in spherical symmetry one may take the Kodama vector\cite{H2} 
as the choice of time determining the preferred vacuum state.
The Kodama vector has vanishing norm at trapping horizons,
becoming zero or null for static wormholes and black holes respectively. 

A more prosaic counter-argument runs as follows.
Quantum field theory in curved space-time 
generally predicts creation of negative-energy particles,
the classic example being Hawking radiation.
For illustrative purposes, 
suppose that all the matter in the universe 
collects into a single black hole with Schwarzschild exterior.
The semi-classical Hawking evaporation of this black hole
will be extremely slow, due to its low Hawking temperature.
However, after an unimaginably long time,
a significant fraction of the energy of the universe 
will have been radiated away, 
with a correspondingly huge amount of negative energy 
having been produced by quantum fluctuations and absorbed by the black hole.
This grossly contradicts 
a naive application of flat-space quantum inequalities to curved space-times.
Recalling that such an evaporating black hole is also a dynamic wormhole, 
any such objection to wormholes would anyway contradict 
the accepted nature of Hawking radiation.

Finally, it is useful for theoretical purposes to have a simple matter model 
which allows negative energy.
One such suggestion is a Klein-Gordon field 
whose gravitational coupling takes the opposite sign to normal.
That is, the Lagrangian and therefore energy tensor take the new sign,
while the Klein-Gordon equation itself is therefore preserved.
This may be thought of as a simple model of negative-energy matter or radiation,
respectively in the massive and massless cases.
A nice coincidence is that 
the massless field is a source for a wormhole of Morris \& Thorne 
which they considered simple enough for an examination question 
for a first course in general relativity.
The dynamical behaviour of this wormhole under perturbations 
is under investigation.

\medskip
Research supported by a European Union Science and Technology Fellowship.
Thanks to Matt Visser for sharing thoughts on how to define dynamic wormholes,
and to Tom Roman, Larry Ford, Chris Fewster and Edward Teo 
for discussions on quantum inequalities.


\begin{references}
\bibitem{MT}M S Morris \& K S Thorne, {Am. J. Phys.} {\bf 56}, 395 (1988).
\bibitem{HV1}D Hochberg \& M Visser, {Phys. Rev.} {\bf D58}, 044021 (1998).
\bibitem{HV2}D Hochberg \& M Visser, {Phys. Rev. Lett.} {\bf 81}, 746 (1998).
\bibitem{HE}S W Hawking \& G F R Ellis, 
The large scale structure of space-time, Cambridge University Press (1973).
\bibitem{H1}S A Hayward, {Phys. Rev.} {\bf D49}, 6467 (1994).
\bibitem{H2}S A Hayward, {Class. Quantum Grav.} {\bf 15}, 3147 (1998).
\bibitem{H3}S A Hayward, Wormhole dynamics (in preparation).
\bibitem{KH}M Kriele \& S A Hayward, {J. Math. Phys.} {\bf 38}, 1593 (1997).
\bibitem{FSW}J L Friedman, K Schleich \& D M Witt, {Phys. Rev. Lett.} 
{\bf 71}, 1486 (1993).
\bibitem{MH}S Mukohyama \& S A Hayward, 
Quasi-local first law of black-hole dynamics (in preparation).
\bibitem{HPS}D Hochberg, A Popov \& S V Sushkov, {Phys. Rev. Lett.} 
{\bf 78}, 2050 (1997).
\bibitem{HK}D Hochberg \& T W Kephart, {Phys. Rev.}
{\bf D47}, 1465 (1993).
\bibitem{FR1}L H Ford \& T A Roman, {Phys. Rev.} {\bf D53}, 5496 (1996).
\bibitem{FR2}L H Ford \& T A Roman, {Phys. Rev.} {\bf D55}, 2082 (1997).
\bibitem{PF}M J Pfenning \& L H Ford, {Phys. Rev.} {\bf D57}, 3489 (1998).
\bibitem{FT}C J Fewster \& E Teo, 
Bounds on negative energy densities in static space-times (gr-qc/9812032).
\bibitem{FPR}L H Ford, M J Pfenning \& T A Roman, {Phys. Rev.} {\bf D57}, 
4839 (1998).
\end{references}
\end{document}